\documentclass{eas}
\usepackage{graphicx}
\begin{document}

\title{ What do stars tell us about planets? 
Asteroseismology of exoplanet-host stars } 
\runningtitle{Asteroseismology of exoplanet-host stars}
\author{ Sylvie Vauclair }\address{ Laboratoire d'Astrophysique de Toulouse et Tarbes ; CNRS ; Universit\'e de Toulouse ; Institut universitaire de France }
\begin{abstract}
Studying the internal structure of exoplanet-host stars compared to that of similar stars without detected planets is particularly important for the understanding of planetary formation. The observed average overmetallicity of stars with planets is an interesting point in that respect. In this framework, asteroseismic studies represent an excellent tool to determine the structural differences between stars with and without detected planets. It also leads to more precise values of the stellar parameters like mass, gravity, effective temperature, than those obtained from spectroscopy alone. Interestingly enough, the detection of stellar oscillations is obtained with the same instruments as used for the discovery of exoplanets, both from the ground and from space. The time scales however are very different, as the oscillations of solar type stars have periods around five to ten minutes, while the exoplanets orbits may go from a few days up to many years.  Here I discuss the asteroseismology of exoplanet-host stars, with a few examples.
\end{abstract}
\maketitle
\section{ Why bothering about asteroseismology while studying planets?}

I can see at least four different reasons for which astrophysicists interested in exoplanets should also bother about the asteroseismology of the central stars of planetary systems.

First reason: the observations for stellar oscillations and exoplanet searches are done with the same instruments. In some cases, the same observations, analysed on different time scales, can lead to both planet detection and seismic studies. This was the case for the star $\mu$ Arae, observed with HARPS during eight nights in June 2004: these observations, aimed for asteroseismology, lead to the discovery of the exoplanet $\mu$ Arae d (Santos et al. \cite{santos04}). 

Second reason : "Some people's noise is other people's signal". Indeed, when searching for exoplanets, the signal to noise ratio is limited by the stellar oscillations, which appear as a noise for the radial velocity variations induced by the planetary motions, while they represent in fact the stellar oscillations signal. A better knowledge of seismology and a better treatment of this low time scale signal could help determining the planetary parameters.

Third reason: obtain precise values of the parameters of exoplanet-hosts stars. Asteroseismic studies, combined with spectroscopic observations, can lead to values of the stellar parameters which are much more precise than from spectroscopy alone. In the case of $\iota$ Hor, for example, the mass determined from spectroscopy and position in the HR diagram was $10\%$ wrong (Vauclair et al. \cite{vauclair08}).

Fourth reason: links between asteroseismology and planets discoveries. In two cases at least, asteroseismic studies lead to exoplanet discoveries. One is the case, already mentioned, of $\mu$ Arae. Studies of seismic period variations (the so-called ``time delay method") also lead recently to the spectacular discovery of a planet around the extreme horizontal branch star V391 Peg (Silvotti et al. \cite{silvotti07}).

Generally speaking, the goals of the asteroseismology of exoplanets-host stars are to derive their masses, ages, evolutionary stages, outside and inside metallicities, to compare them with the Sun and with stars without detected planets, to obtain hints about the theories of planetary formation and migration, and to try and model the oscillations well enough to decrease the "noise" for planet detection.

\section{ Basics of asteroseismology for slowly rotating solar-type stars }

In solar-type stars, acoustic waves are permanently created by the motions which occur in their outer layers, induced by convection and related processes. The waves are damped, but as other waves always appear, the stellar sphere globally behaves as a resonant cavity, and the oscillations can be treated, with a very good approximation, as standing waves. 

Each mode can be characterized with three numbers: the number of nodes in the radial direction, $n$, and the two tangential numbers $\ell$ and $m$, which appear in the development of the waves on the spherical harmonics:

\begin{equation}
Y_l^m (\theta,\phi) = (-1)^m C_{l,m} P_l^m (cos\theta) exp(im\theta)
\end{equation}

Several combinations of the oscillation frequencies are used to obtain more precise constraints on the stellar internal structure, like the ``large separations", which are defined as the differences between two consecutive modes of the same $\ell$ number, and the ``small separations", defined as:

\begin{equation}
\delta\nu = \delta \nu_{n,l} - \nu_{n-1,l+2}
\end{equation}

For acoustic modes in a given star, the large separations are nearly constant, and their average value is about equal to half the inverse of the stellar acoustic time, i.e. the time needed for the $l = 0$ waves to travel along the whole radius. 

Meanwhile, the small separations present variations which are directly related to the structure of the stellar cores (Roxburgh \& Vorontsov \cite{roxburgh94}, Roxburgh \cite{roxburgh07}, Soriano et al. \cite{soriano07}, Soriano \& Vauclair \cite{soriano08}).

An interesting tool used to compare stellar models with the observations is the ``echelle diagram". This diagram presents the mode frequencies in ordinates, and the same frequencies modulo the large separations in abscissae (see Figure 2).

\section{ Asteroseismology of exoplanet-host stars }

Aseroseismology of exoplanet-host stars is a useful tool to determine their internal structure and their behaviour with respect to stars without detected planets. Let us recall that, due to the present observation bias, many stars may very well host undetected planets. The presently detected planets have to be close to the stars, and their orbital plane must not be perpendicular to the line of sight. In this framework, the special characteristics of exoplanet-host stars compared to the other ones is more related to the migration process of the planets than to their mere existence.

An important particularity of exoplanet-host stars is their overmetallicity, compared to the other stars and to the Sun (e.g. Santos et al. \cite{santos05}). The accretion origin for this overmetallicity (planet engulfment at the beginning of the stellar system formation) is now ruled out for several reasons. One of these reasons concerns the enormous amount of metallic matter which should be accreted to explain the observed overabundances. Furthermore, the accreted metals would not stay in the outer convective zones of the stars, but would fall down due to thermohaline convection induced by the unstable inverse $\mu$-gradient (Vauclair \cite{vauclair04}). The seismic analysis of the star $\iota$ Hor (see below) also proves that the overmetallicity was there at the origin, in the cloud out of which the stellar system formed. 

A crucial unknown parameter, very important for the determination of the stellar characteristics, is the helium abundance. Unfortunately, helium is not directly observable in the spectra of solar-type stars.
If the stellar systems form inside an overmetallic interstellar cloud, is this cloud also helium-enriched as predicted from theories of the chemical evolution of galaxies or not? This depends on the stellar mass function and there is no clear answer at the present time. Detailed seismic analysis can solve this question, as has been proved for $\iota$ Hor: in this star, the helium abundance is not larger than solar, it may even be smaller (Vauclair et al. \cite{vauclair08}).

Up to now four exoplanet-host stars have been observed on relatively long periods (8 or 9 consecutive nights) with the HARPS and SOPHIE spectrometers. They are: $\mu$ Arae (HARPS, 2004), $\iota$ Hor (HARPS, 2006), 51 Peg (SOPHIE, 2007) and 94 Cet (HARPS, 2007). Figure 1 presents, as an example, part of the radial velocity oscillation curves obtained for 51 Peg (Soriano et al., in preparation). Several other exoplanet-host solar type stars have been observed for about half an hour: they all oscillate, without any exception. 

At the present time, complete modelling has been performed for $\mu$ Arae and $\iota$ Hor only. 
The models have been computed using the TGEC (Toulouse-Geneva stellar evolution code), with the OPAL equation of state and opacities (Rogers \& Nayfonov \cite{rogers02}, Iglesias \& Rogers \cite{iglesias96}) and the NACRE nuclear reaction rates (Angulo et al. \cite{angulo99}). Microscopic diffusion is included in all the models (Paquette et al. \cite{paquette86}, Richard et al. \cite{richard04}). The treatment of convection is done in the framework of the mixing length theory and the mixing length parameter is adjusted as in the Sun. Adiabatic oscillations frequencies are computed using the adiabatic PULSE code (Brassard \cite{brassard92}).

The only COROT main target hosting an exoplanet, HD52265 has also been modelled in advance before the observations, which should be done during the fall 2008 (Soriano et al. \cite{soriano07}). I give below some information about $\mu$ Arae and $\iota$ Hor

\begin{figure*}
\begin{center}
\includegraphics[width=12cm]{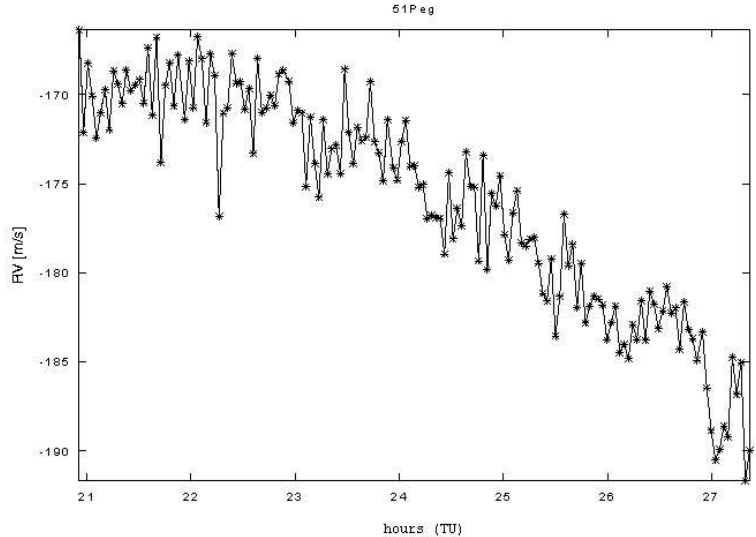}
\end{center}
\caption{The star 51 Peg has been observed for asteroseismology with the SOPHIE spectrometer, at Haute Provence Observatory, in August 2007. The comparisons with models are still in progress (Soriano et al., in preparation). The graph presents part of the radial velocity curve, obtained during one night. The solar-like oscillations are clearly visible. The overall decrease of the radial velocity during the night is the signature of the exoplanet discovered by Mayor and Queloz \cite{mayor95}}
\label{fig1}
\end{figure*}

\section{ The saga of $\mu$ Arae }

The exoplanet-host star $\mu$ Arae (HD160691, HR6585, GJ691) is a G5V star with a visual magnitude V~=~5.1, an Hipparcos parallax $\pi$~=~65.5~$\pm$~0.8 mas, which gives a distance to the Sun of 15.3 pc and a luminosity of $\log L/L_{\odot}~=~$0.28$~\pm~$0.012. This star was observed for seismology in August 2004 with HARPS. At that time, two planets were known. The observations aimed for seismology lead to the discovery of a third planet, $\mu$ Ara d, with period 9.5 days (Santos et al. \cite{santos04}). More recently, evidence for a fourth planet has been discovered (Pepe et al. \cite{pepe07}).

The HARPS seismic observations allowed to identify 43 oscillation modes of degrees $l~=~0$ to $l~=~3$ (Bouchy et al. 2005). In Figure 2, they are presented in the form of an echelle diagram and compared with a model. From the analysis of the frequencies and comparison with models, the values T$_{eff}$~=~5770~$\pm$~50~K and [Fe/H]~=~0.32~$\pm$~0.05~dex were derived. Spectroscopic observations by various authors gave five different effective temperatures and metallicities (see references in Bazot et al. \cite{bazot05}). The values obtained from seismology are much more precise than those obtained from spectroscopy alone.

\begin{figure*}
\begin{center}
\includegraphics[totalheight=8cm,width=10cm]{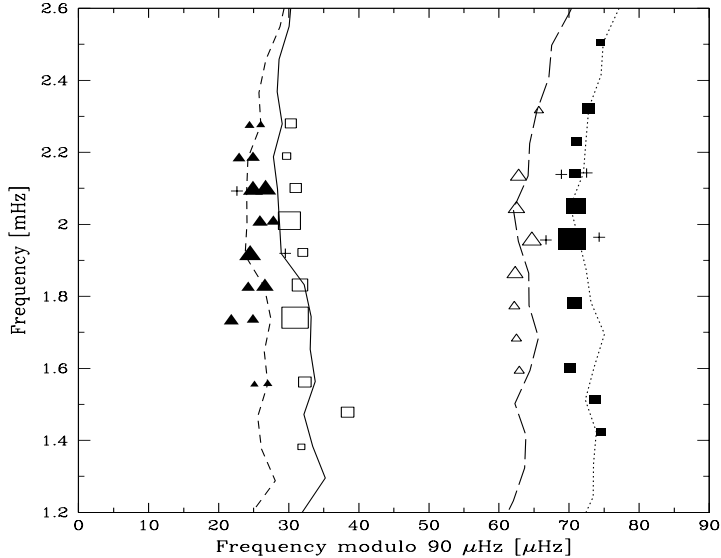}
\end{center}
\caption{ Echelle diagram for the star $\mu$ Arae. The ordinate represents the frequencies of the modes and the absissa the same frequencies modulo the large separation, here 90~$\mu$Hz. The symbols represent the observed oscillation frequencies. The lines represent the results of the computations for one of the best model, obtained with T$_{eff}$~=~5770~$\pm~$50~K; [Fe/H]~=~0.32~$\pm$~0.05~dex; $M~=~1.18~\pm~0.01~M_{\odot}$; age~4.25~$\pm$~0.1~Gyr. The four lines correspond respectively, from left to right, to $\ell$~=~2, 0, 3 and 1. The symbols for the observed frequencies have been chosen according to their mode identifications (after Bazot et al. \cite{bazot05}).}
\label{fig2}
\end{figure*}

\subsection{ The special case of $\iota$ Hor }

Among exoplanet-host stars, $\iota$ Hor is a special case for several reasons (see Laymand \& Vauclair \cite{laymand07} and Vauclair et al. \cite{vauclair08}). Three different groups have given different stellar 
parameters for this star: Gonzalez et al. \cite{gonzalez01}, Santos et al. \cite{santos04} and Fischer \& Valenti \cite{fischer05}. Meanwhile, Santos et al. \cite{santos04} suggested a mass of 1.32 M$_{\odot}$ while Fischer and Valenti \cite{fischer05} gave 1.17 M$_{\odot}$.

Some authors (Chereul et al. \cite{chereul99}, Grenon \cite{grenon00}, Chereul and Grenon \cite{chereul00}, Kalas and Delorn \cite{kalas06}, Montes et al. \cite{montes01}) pointed out that this star has the same kinematical characteristics as the Hyades: its proper motion points towards the cluster convergent. Two different reasons were possible for this behaviour: either the star formed together with the Hyades, in a region between the Sun and the centre of the Galaxy, which would explain its overmetallicity compared to that of the Sun, or it was dynamically canalized by chance (see Famaey et al. \cite{famaey07}).

Solar-type oscillations of $\iota$ Hor were detected with HARPS in November 2006. Up to 25 oscillation modes could be identified and compared with stellar models. The results lead to the following conclusions for $\iota$ Hor (Vauclair et al. \cite{vauclair08}): (Fe/H) is between 0.14 and 0.18; the helium abundance Y is small, 0.255 $\pm $ 0.015; the age of the star is 625 $\pm$ 5 Myr; the logarithm of the gravity is 4.40 $\pm$ 0.01 and its mass $1.25 \pm 0.01 M_{\odot}$. The values obtained for the metallicity, helium abundance and age of this star are those characteristic of the Hyades cluster (Lebreton et al. \cite{lebreton01}).

In summary, we have found from seismic analysis that this exoplanet-host star has been formed together with the Hyades cluster. As a consequence, the overmetallicity has been present from the beginning and is not due to accretion. The stellar mass has also been derived with a much better precision than from spectroscopy alone. It lies in between the two different values given by Santos et al. \cite{santos04} and Fischer \& Valenti \cite{fischer05}.

\section{ Conclusion }

From the few examples already available, asteroseismology has proved to be a powerful tool for determining stellar parameters, in particular for exoplanet-host stars. The scientific community is well prepared for future observations with on-going or planned projects. Space missions like COROT, and later on KEPLER, are expected to give a large amount of new data for seismology, besides planet searches. Meanwhile ground based instruments devoted to exoplanets like HARPS or SOPHIE can be used for seismology. All these observations will give the possibility of using the seismic tests with a precision never obtained before. They will lead to better parameters for exoplanet-host stars and help for a better understanding of planetary formation and evolution.


\end{document}